# Terahertz bandwidth integrated radio frequency spectrum analyzer via nonlinear optics

**Marcello Ferrera,[1,2,*] Christian Reimer,[1] Alessia Pasquazi,[3]
Marco Peccianti,[3] Matteo Clerici,[1,2] Lucia Caspani,[1] Sai T. Chu,[4] Brent E. Little,[5]
Roberto Morandotti,[1] and David J. Moss[1,6]**

[1]*INRS-EMT, 1650 Boulevard Lionel Boulet, Varennes, Québec J3X 1S2, Canada*
[2]*School of Engineering and Physical Sciences, Heriot-Watt University,
David Brewster Building, Edinburgh, Scotland EH14 4AS, United Kingdom*
[3]*Department of Physics and Astronomy, University of Sussex, Falmer, Brighton BN1 9RH, United Kingdom*
[4]*Department of Physics and Materials Science, City University of Hong Kong, Tat Chee Avenue, Hong Kong, China*
[5]*Xi'an Institute of Optics and Precision Mechanics of CAS, Xi'an, China*
[6]*School of Electrical and Computer Engineering, RMIT University, Melbourne, Victoria 3001, Australia*
[*]*M.Ferrera@hw.ac.uk*

**Abstract:** We report an integrated all-optical radio frequency spectrum analyzer based on a ~ 4cm long doped silica glass waveguide, with a bandwidth greater than 2.5 THz. We use this device to characterize the intensity power spectrum of ultrahigh repetition rate mode-locked lasers at repetition rates up to 400 GHz, and observe dynamic noise related behavior not observable with other techniques.

## 1. Introduction

The growing demand for ultra-fast data transmission and processing is creating a need for devices that function well beyond the speed of electronics. Photonics offers the capability of generating and measuring ultrashort optical pulses, with bandwidths of several THz and at repetition rates of 100's of GHz. Performing temporal diagnostics at these speeds is extremely difficult and yet essential to achieve high optical signal fidelity of fundamental noise parameters such as time jitter and amplitude noise, critical for obtaining the maximum performance of many devices such as high frequency - clock optical modules [1, 2].

The radio frequency (RF) spectrum (or power spectrum) of an optical signal with intensity $I(t)$ is the power spectral density of the autocorrelation of the optical intensity: $P(\omega) \propto \int [I(t) \otimes I(t)] exp(i\omega t) dt$; where $\omega$ is the angular frequency. Through the direct analysis of the RF spectrum, it is possible to perform a full characterization of several fundamental noise parameters of optical waveforms, such as the intensity and pulsewidth fluctuations, the repetition rate, the cavity length mismatch, the time jitter, and more [3].

The traditional way of retrieving the RF spectrum consists of recording the temporal intensity profile by way of an ultra-fast photo detector and then processing this signal either numerically or using electronic circuitry in a double-conversion super-heterodyne scheme [4]. The electronic components limit the bandwidth of this approach to around 100 GHz [5]. Although RF spectrum analysis is broadly used to characterize mode-locked lasers, its limited bandwidth has precluded its use with ultra-high high repetition rate lasers.

Nonlinear integrated photonics has recently provided powerful all-optical tools to overcome the speed limitation associated to electronics [6-11]. A key breakthrough was the scheme for an all-optical RF-spectrum analyzer, introduced by Dorrer and Maywar [12], that exploits optical mixing between a signal under test and a CW probe, via the Kerr ($n_2$) nonlinearity in highly nonlinear fibers or integrated waveguides. In this approach, a single measurement of the optical spectrum $S(\omega) \propto \int [E(t) \otimes E(t)] exp(i\omega t) dt$ around the CW probe, performed by an optical spectrum analyzer (OSA), yields the intensity power spectrum of the signal under test – $P(\omega) \propto \int [I(t) \otimes I(t)] exp(i\omega t) dt$. This strategy can achieve much broader bandwidths than electronic methods, with an intrinsic trade-off between sensitivity and bandwidth, or equivalently between the nonlinear response and total dispersion of the waveguide [13]. For example, increasing the nonlinear device length enhances the nonlinear response, or sensitivity, but results in an increased dispersion that reduces the frequency response. For this reason, an optical integration platform with high nonlinearity and low net dispersion (waveguide plus material) is highly desirable, in order to have the best overall performance in sensitivity and bandwidth.

The first demonstration of an integrated all-optical RF spectrum analyzer [13] was achieved in dispersion engineered chalcogenide waveguides, which provided the required nonlinear response in only a few centimeters of length. This was followed by the first demonstration of an RF spectrum analyzer in a CMOS compatible platform based on a silicon nanowire [14]. While this device achieved comparable performance to that reported in [13], and with a much shorter device length, it was nonetheless limited in bandwidth by a relatively high dispersion and in overall performance by saturation and free carrier effects [15].

It is worth mentioning that all the integrated nonlinear RF spectrum analyzers reported up to date have been focused only on few of the most fundamental requirements (e.g. nonlinearities, dispersion, losses, CMOS compatibility) for maximizing the system performances. Here, for the first time, we report about a system for the RF analysis which addresses all the previously listed requirements at once. Our integrated RF spectrum analyzer [see Fig. 1] is based on a nonlinear CMOS-compatible platform with extremely low linear losses, plus negligible nonlinear saturation and dispersion in the telecom band [16]. Our waveguides were deposited by standard chemical vapour deposition (CVD) and the device patterning and fabrication were performed using photolithography and reactive ion etching to produce exceptionally low sidewall roughness. The final fabrication step consists in over-coating the core device with a silica glass upper cladding. No post-processing annealing was required in order to obtain the ultra-low losses possessed by our waveguide (0.06dB/cm).

We use this device to measure the RF spectrum of a micro-resonator based modelocked laser [17] emitting sub-ps pulses at repetition rates of 200 GHz and 400 GHz. The RF spectra obtained in this manner allows us to analyze these lasers according to the noise burst model, which identifies very rapid intensity fluctuations of the laser pulses as the main source of noise [18-19]. The RF spectra for both lasers show sensitivity to high frequency noise not detectable by other methods and it is of fundamental use to characterize a source which shows great potential especially for applications in the quantum communications domain [20].

## 2. Principle of Operation

The principle of an all-optical RF spectrum analyzer is shown in Fig. 1. A CW source is combined with a signal under test and then directed into a nonlinear medium, the output of which is analyzed by an optical spectrum analyzer. Our device is based on a ~ 4 cm long spiral nonlinear waveguide (cross section $1.5 \times 1.45 \mu m^2$) [21-23].

The general concept of the technique is based on the cross phase modulation (XPM) experienced by the weak monochromatic probe (CW) due to the intensity variation of the signal, centered at an optical frequency $\omega_0$. The CW probe beam, represented by an electric field $E_{CW}(t) = E_0 \exp(-i\omega_0 t)$, propagates inside the $\chi^{(3)}$ nonlinear waveguide along with the signal under test. The signal induces XPM in the phase of the CW probe in proportion to its intensity $I_s(t)$. If this modulation is small, it is easy to show [12] that the CW field $E_{CW}^{out}(t)$ at the output of the waveguide is simply:

$$E_{CW}^{out}(t) = E_{CW}(t)\exp(i\alpha I_s(t)) \approx E_0 \exp(-i\omega_0 t)(1 + i\alpha I_s(t)) \quad (1)$$

Where $\alpha$ is a constant proportional to the waveguide nonlinearity and length. The optical spectrum $P_{CW}^{out}(\omega)$ is easily obtained by Fourier transforming the autocorrelation of Eq. (1):

$$E_{CW}^{out}(t) \otimes E_{CW}^{out}(t) = \lim_{T\to\infty} \frac{1}{T} \int_{-T/2}^{T/2} E_{CW}^{out}(t+t)\left[E_{CW}^{out}(t)\right]^* dt \propto \exp(-iW_0 t)\left(1 + aI_s(t) \otimes aI_s(t)\right) \quad (2)$$

thus obtaining:

$$S_{CW}^{out}(\omega) \propto \delta(\omega - \omega_0) + \alpha^2 P_s(\omega - \omega_0) \quad (3)$$

where $P_s(\omega)$ is the intensity power spectrum, i.e. the Fourier transform of the autocorrelation of the optical intensity $I_s(t)$. The optical spectrum, $S_{CW}^{out}(\omega)$, at the output of the nonlinear waveguide can be written as the sum of a Dirac delta function (the original spectrum of the CW probe) and a term which is proportional to the RF spectrum of the signal ($P_s(\omega)$), both of which are centered at $\omega_0$.

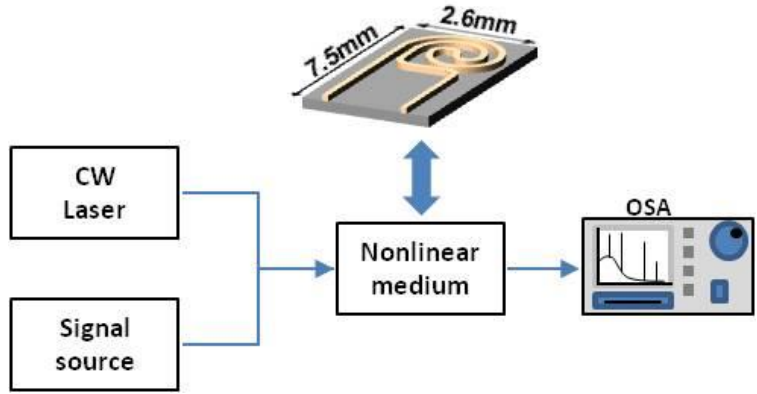

Fig. 1. General schematic of the nonlinear RF spectrum analyzer. The pulsed signal under test is combined with a CW laser and they interact in a nonlinear medium, in this case a 4cm-long doped silica-glass spiral waveguide. The signal after the nonlinear interaction is detected using a standard optical spectrum analyzer (OSA).

## 3. Frequency Response

The bandwidth of an RF spectrum analyzer is probably its most important characteristic, and since the bandwidth of all-optical RF spectrum analyzers is limited only by the nonlinear interaction [12], they operate at frequencies far beyond the capability of electronics. Hence they must be characterized optically, and the most common approach is an optical heterodyne method where two CW waves are interfered together to generate an RF tone at the difference frequency. The two CW probes are centered at $\omega_S^\pm = \omega_S^0 \pm \Delta\omega_S/2$ (where $\Delta\omega_S$ is the frequency difference between the two probes and $\omega_S^0$ is the signal central frequency), and when interfered with each other, they generate an RF beat frequency tone at $\Delta\omega_S$. The resulting modulated beam then induces XPM on the CW probe beam, with sidebands at $\omega_0^\pm$ given by:

$$\omega_0^+ = \omega_0 + \omega_S^+ - \omega_S^- = \omega_0 + \Delta\omega_S \qquad (4)$$

$$\omega_0^- = \omega_0 + \omega_S^- - \omega_S^+ = \omega_0 - \Delta\omega_S \qquad (5)$$

The 3dB bandwidth of the device is then determined by measuring the XPM induced peak on the probe beam as the beat frequency of the signal is swept.

A critical quantity is the difference in phase velocity, or propagation constants ($\beta(\omega)$), of the different beams in the nonlinear waveguide:

$$\Delta\beta^+ \equiv \beta(\omega_0^+) - \beta(\omega_0) - \beta(\omega_S^+) + \beta(\omega_S^-) \approx \left(\beta^{(1)}(\omega_0) - \beta^{(1)}(\omega_S)\right)\Delta\omega_S + \frac{1}{2}\beta^{(2)}(\omega_0)\Delta\omega_S^2 \qquad (6)$$

$$\Delta\beta^- \equiv \beta(\omega_0^-) - \beta(\omega_0) - \beta(\omega_S^-) + \beta(\omega_S^+) \approx -\left(\beta^{(1)}(\omega_0) - \beta^{(1)}(\omega_S)\right)\Delta\omega_S + \frac{1}{2}\beta^{(2)}(\omega_0)\Delta\omega_S^2 \qquad (7)$$

where $\beta^{(1)}$ and $\beta^{(2)}$ represent the first- and second-order derivative, respectively. The RF bandwidth of the device is limited by the group velocity "walk-off", or mismatch, between the signal and the CW probe, which degrades the efficiency of the XPM process. Thus, minimizing dispersion directly increases the bandwidth, and the device we report here has very low dispersion in the C-band [22].

Figure 2 shows the experimental setup for measuring the frequency response of the device. Two tunable narrow linewidth lasers, labeled CW-1 and CW-2, offset in frequency, were combined together to generate the signal while a third laser (CW-3) acted as a probe beam. Amplifiers were used to compensate for optical loss (largely from the couplers), while polarization controllers ensured the same polarization (TM) for all three CW beams coupled into the waveguide.

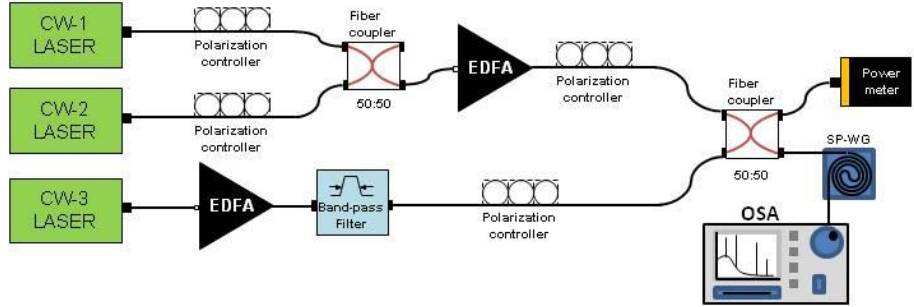

Fig. 2. Experimental setup for the estimation of the all-optical RF spectrum analyzer operative bandwidth. Four polarization controllers ensure the same polarization (TM) for all the three CW signals coupled into the doped silica glass waveguide. The use of a band-pass filter was necessary to filter out the noise generated by one of the two amplifiers. The overall power at the waveguide input was constantly monitored by a power meter while the optical spectra were recorded by means of an OSA.

A band-pass filter consistently reduces the noise due the amplified stimulated emission (ASE) from the probe amplifier. The power at the waveguide input was constantly monitored by a power meter while the optical spectra were recorded by an OSA. The measurements [see Fig.

3(a)] were done by fixing the wavelengths of two lasers - CW-1 at $\lambda_1 \approx 1550$nm and CW-3 at $\lambda_3 \approx 1590$nm, respectively. In order to vary the frequency of the RF tone we swept $\lambda_2$ (CW-2) between $\lambda_1$ and $\lambda_3$, while recording the optical spectrum over a broad wavelength range around $\lambda_3$. The nonlinear waveguide generated a new peak via XPM between the modulated signal and probe, at a wavelength $\lambda_4 > \lambda_3$ [see inset Fig. 3(a)], approximately equidistant from $\lambda_3$ as $\lambda_2$ is from $\lambda_1$ (the spacing is identical in frequency - see Eq. (4) and (5)).

The amplitude of the peak at $\lambda_4$ diminishes as $\lambda_2$ is tuned away from $\lambda_1$ since this increases the frequency of the RF tone, making it more susceptible to dispersion induced walk-off delay between the signal and the probe. The experimentally measured bandwidth – i.e., the 3dB reduction in the peak power of $\lambda_4$ (1$^{st}$ RF component) is limited by the intrinsic device bandwidth of the nonlinear process, including the walk-off effects previously discussed, and/or by experimental limitations due to spurious nonlinear effects or ASE noise from the erbium doped fiber amplifiers (EDFAs) (particularly the high power EDFA used for CW-3). In our case the measurements were also limited by the operation bandwidth (C-band) of our lasers and amplifiers.

The results are shown in Fig. 3(b), where the normalized peak power at $\lambda_4$ is plotted as a function of the detuning between $\lambda_1$ and $\lambda_2$. Since the operative bandwidth depends on numerous parameters not all of them quantitatively defined in our analysis the experimental data fitting is performed by using a sigmoid function. The root mean square of the error (Standard Deviation) is equal to 0.00676 while the average error bar [reported in Fig. 3(b)] is <2%. As expected, $\lambda_4$ precisely tracked the wavelength detuning between $\lambda_1$ and $\lambda_2$, and we measured a 3dB bandwidth ($f_{3dB}$) of about 20nm, corresponding to 2.5 THz. We believe that this value was primarily limited not by the intrinsic response of the device since simple estimates of this are extremely promising (see below) but by our experimental measurement capability. One factor in our measurements was the onset of unwanted four-wave mixing (FWM) processes such as the semi-degenerate FWM between $\lambda_2$ and $\lambda_3$ that resulted from the limited tuning range (C-band) of our lasers. As the detuning between $\lambda_1$ and $\lambda_2$ was increased, $\lambda_2$ approached $\lambda_3$, enhancing new and unwanted phasematched processes not related to the signal RF spectrum.

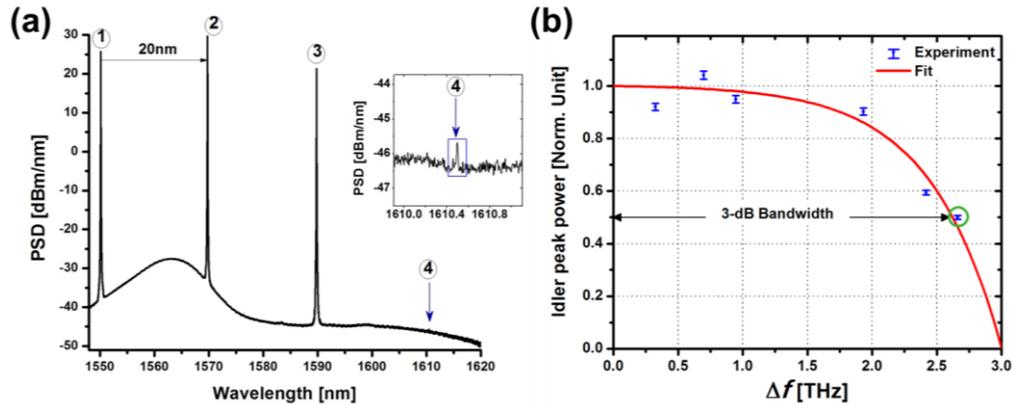

Fig. 3. Bandwidth measurement of our RF spectrum analyzer. a) Power spectral density for λ1↔λ2=20nm. In the inset we zoom on the idler peak. b) Normalized idler peak power as a function of the frequency detuning between signal 1 and 2. The green circle indicates that for a detuning exceeding 2.5 THz spurious nonlinear components start being visible.

The data point highlighted by the green circle, at $\Delta f \sim 2.7$THz in Fig. 3(b), corresponds to the condition at which spurious nonlinear processes start being visible. In other words, whatever competing nonlinear process (which is not the XPM between the signal and the seed) producing spectral components inside the frequency support of the RF spectrum has to be considered highly detrimental for the overall RF analysis. In fact, these spurious spectral lines would be misinterpreted as RF noise of the original signal while they were not.

It is important to keep in mind that the operational bandwidth of the system does not solely depend on the nature of the waveguide dispersion curve. The nonlinear gain curve together with the wavelength location of both the signal and the seed are also very important parameters to determine the operational bandwidth. In our case, the spurious components could be reduced or eliminated by shifting both CW lasers comprising the signal well out of the C-band, but not only was this outside the capability of our C-band components, it would in fact have decreased the intrinsic bandwidth of the device (by increasing the signal/probe walk-off). It is worth mentioning that the remarkable operational bandwidth of 2.5THz has been obtained without the use of any dispersion engineering process. It is reasonable to believe that such optimization process could consistently improve the device performances [24]

A simple estimate [13] of our intrinsic device bandwidth due to walk-off yields $f_{3dB}$ = 130.2 THz (see subsequent "Discussion" paragraph) - a result of the extremely low dispersion of our device. While we expect that this would be reduced by higher order dispersion or other effects, we nonetheless believe that the intrinsic response of our device is likely much higher than the experimentally measured $f_{3dB}$ = 2.5THz – possibly on the order of 10's of THz. This would make its bandwidth truly unprecedented and render it applicable to a wide range of new applications such as monitoring of extremely high bandwidth superchannels [25].

## 4. Characterization of an ultrahigh repetition rate mode-locked laser

A periodic pulsed source, such as a mode locked laser with a slowly varying envelope $F(\omega)$ with pulse duration $t_p$ at a repetition rate of $T$, has a spectral bandwidth inversely proportional to $t_p$. For a non-ideal laser affected by amplitude noise and time jitter, the RF spectrum $P_S(\omega)$ is [3]:

$$P_S(\omega) = F(\omega)\left[\sum_{k=0}^{\infty}\delta\left(\omega-\frac{2k\pi}{T}\right) + P_k^A\left(\omega-\frac{2k\pi}{T}\right) + (2k\pi)^2 \cdot P_k^j\left(\omega-\frac{2k\pi}{T}\right)\right] \qquad (8)$$

where the slowly varying envelope $F(\omega)$ is "sampled" in frequency by a train of Dirac delta functions arising from the train of perfect noise-free pulses.

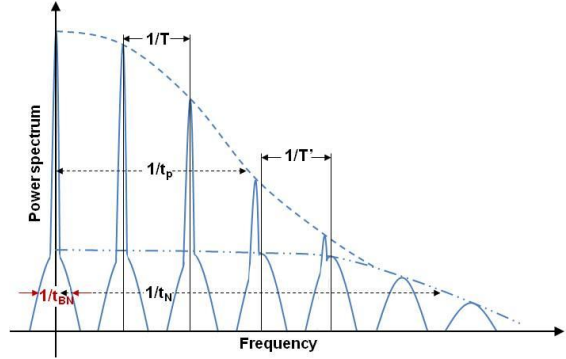

Fig. 4. Typical RF spectrum $P_S(\omega)$ of a non-ideal mode-locked laser source affected by amplitude noise and time jitter. All the standard parameters evaluated from the RF spectrum, e.g. transform-limited duration of the optical pulse ($t_p$), repetition rate of the source ($1/T$), reciprocal of the cavity round trip time ($1/T'$), noise band of random fluctuation in the pulse train, and the reciprocal of the duration of such a fluctuations ($1/t_N$), are also indicated.

The other terms account for the non–ideality of the signal: $P_k^A(\omega)$ and $P_k^J(\omega)$ are the power spectra related to the amplitude noise and random temporal jitter, respectively. A typical experimental RF power spectrum looks like that shown in Fig. 4, where one can identify many of the fundamental parameters introduced in Eq. (8): i) the reciprocal of the time duration of the optical pulses ($1/t_p$); ii) the repetition rate of the source ($1/T$); iii) the reciprocal of the cavity round trip time ($1/T'$) - fundamental to evaluate the cavity length mismatch; iv) the noise band of random fluctuations in the pulse train; and v) the reciprocal of the duration of these fluctuations ($1/t_N$).

An RF spectrum such as that shown in Fig. 4, where $1/t_N \gg 1/t_p$, describes the source according to the noise burst model [18,19] where the main source of noise comes from very fast and random fluctuations in the laser intensity. The ability of all-optical RF spectrum analyzers to measure RF spectra over bandwidths of several terahertz is an unprecedented and powerful tool to characterize these extremely fast, short pulsed mode-locked lasers. It will enable fundamental improvements and effective control over future ultra-fast and high-repetition rate optical pulse sources.

We used our all-optical RF spectrum analyzer to characterize a 200GHz mode-locked filter-driven four-wave mixing laser [17, 21], whose internal layout is shown in Fig. 5. This architecture was proposed in order to achieve extremely stable operation at high repetition rates, while maintaining very narrow linewidths for the frequency components. The system is based on a nonlinear monolithic high-Q (quality) factor resonator, which acts as both a linear filter and a nonlinear element. The resonator is embedded in a main fiber cavity loop that contains an erbium-doped fiber amplifier (EDFA) in order to provide gain, as well as a band pass filter and polarization controller. The laser emitted 730fs pulses at a repetition rate of 200 GHz, set by the free spectral range (FSR) of the micro-cavity.

While in [17] both stable and unstable operation were reported, here we used only the unstable configuration characterized by a higher phase and amplitude noise, as well as by the cavity length mismatch. Cavity length mismatch noise is typical of mode-locked lasers that employ a double cavity scheme [3], where noise affects the two cavities involved differently, resulting in a frequency off-set between the maxima of the rapidly varying functions ($P_k^A(\omega)$ and $P_k^J(\omega)$) previously defined.

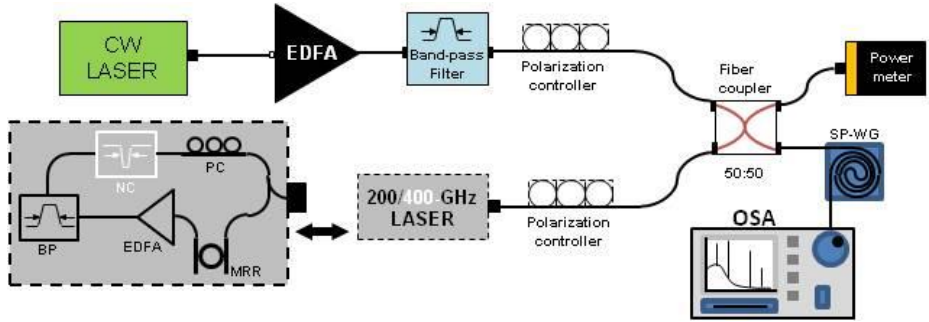

Fig. 5. Experimental setup for measuring the RF spectrum of the modelocked lasers. The source under investigation is a mode-locked filter-driven four-wave mixing laser (grey inset with dashed contours). A CW laser source was first amplified and filtered then properly polarized (TM) before being sent into a 50:50 fiber coupler (top arm). On the second arm (bottom) a mode-locked laser generated a train of optical pulses that served as the signal under test. The signal was TM-polarized before being coupled into the second arm of the fiber coupler. The two outputs of the fiber coupler were directed to the 4cm-long doped silica glass waveguide and a power meter, respectively. Finally, the optical spectrum at the output of the waveguide was recorded by an OSA.

The unstable configuration of our laser, where a large number of modes of the main cavity oscillate inside each resonator line, exhibits frequency offsets between the various peaks in the RF spectrum. In the present work we also studied a modified version of the mode-locked laser that operates at 400 GHz, achieved by introducing a -30dB notch filter (0.8nm/100GHz bandwidth - shown in white in Fig. 5) in the main cavity loop, centered at

λ=1556.15 nm, corresponding to the central frequency of the ring comb. In this way, we effectively doubled the FSR of the system, forcing the laser to oscillate at 400 GHz. The experimental setup [see Fig. 5] consists of two components - the modelocked laser to be studied and the RF spectrum analyzer. The CW probe for the spectrum analyzer was amplified by an EDFA and then filtered by a band-pass filter to remove ASE noise. The amplified beam was then passed through a polarization controller set to TM polarization, followed by a 50:50 fiber coupler. In the other arm, the modelocked laser was also passed through a polarization controller to match the polarization of the CW probe, and then combined with the probe in a 50-50 fiber coupler. One output of the coupler served as a reference and the other was directed to the nonlinear waveguide. The total optical power at the waveguide input was 150 mW, of which 100mW was from the signal and 50mW from the CW probe. The power spectrum at the output of the integrated device was then recorded by an OSA.

In order to maximize the bandwidth of the RF spectrum analyzer, it is optimal for the signal and probe wavelengths to be symmetrically located with respect to the zero dispersion wavelength (frequency) of the waveguide [12], which in our case is near 1560 nm for the TM modes [22]. For this reason, the modelocked laser (signal) was tuned around $\lambda_s$=1550nm, while the probe was located at $\lambda_0$=1570nm.

## 5. Results

The optical spectra for both the 200 GHz and the 400 GHz lasers are shown in Fig. 6, where only the right halves of the spectra centered around the probe frequency are shown for clarity. In fact, the RF spectrum can be obtained from only one half of the optical spectrum since a component of the RF spectrum at frequency $\Omega$ occurs in the optical spectrum at both $\lambda_0+\Omega$ and $\lambda_0-\Omega$. In principle, the other half can be used to improve precision or signal to noise through numerical averaging.

The repetition rate is immediately evident in Fig. 6, from the spacing of the different frequency components. Furthermore, a highly resolved analysis of the frequency profile of our RF spectrum reveals an asymmetric "bump" (indicated by the magenta arrow in the inset of Fig. 6) which we believe to be related to the cavity length mismatch. This distortion in the peaks arises from noise that is typical of mode-locked lasers that employ a double cavity scheme, such as the one used here [3], since the oscillating modes in these lasers must match resonant conditions in both cavities (main loop and micro-resonator) simultaneously. The significantly different environmental and structural nature of the two cavities (micro-resonator versus main cavity) resulted in widely differing random variations in terms of effective lengths, or between 1/T and 1/T'. This manifests in the RF spectrum as a frequency offset between the maxima of the rapidly varying function shown in Fig. 4. In our case, this feature became visible in the first two peaks of the RF spectrum for the 200 GHz signal but only for one peak for the 400 GHz signal. However, because of the limited resolution of the OSA at our disposal (10 pm, 1.25 GHz) and the limited number of RF spectral lines (i.e. the "bump" is apparent only in one peak for the signal at 400GHz and in two for the signal at 200GHz) we were not able to conclusively state that the "bumps" were solely or mainly related to the cavity-length mismatch.

In order to have a detailed and quantitative analysis of the noise component related with the cavity length mismatch a more resolved OSA is needed. OSAs with a resolution of 40 fm (5 MHz) (APEX Tech.) are commercially available. A less expensive strategy could be that one of adopting a high resolution configuration for the spectral analysis. One example is the one based on stimulated Brillouin scattering (SBS) amplification which has already been proposed and demonstrated with resolution of few tens of MHz. [26, 27].

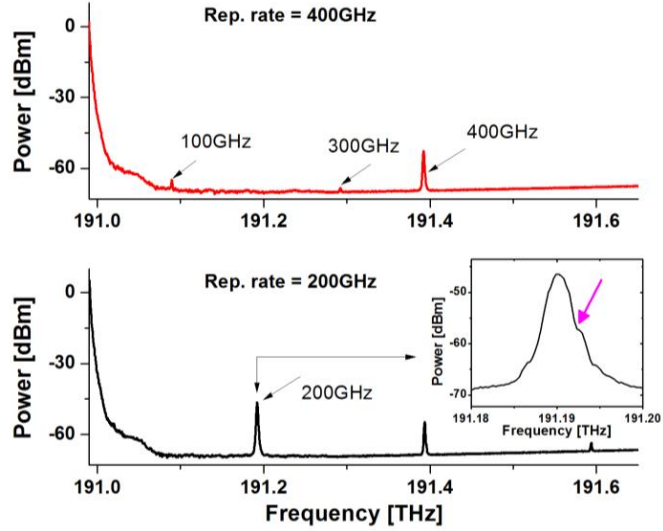

Fig. 6. Experimental RF spectra as seen from the optical spectra of both lasers - 200 GHz (bottom) and 400 GHz (top) near the probe frequency. The inset is a zoom of the first RF component for the signal at 200 GHz. The magenta arrow (color online) in the inset points at the typical feature related to the cavity length mismatch noise. The same feature is seen in the second components of the RF spectrum at 200 GHz and the first component of the signal at 400 GHz.

From the RF spectrum of the signal at 200GHz, we roughly estimate the minimal time duration of laser pulses to be ≈ 840fs - only 15% larger than that measured via autocorrelation. As evident from the RF spectrum of the 400 GHz signal, we also detected a 100 GHz noise component (undetectable in the optical spectrum of the laser source), arising from the introduction of the 0.8nm (100GHz) notch filter. These noise peaks (marked 100 GHz and 300 GHz in Fig. 6) come from spurious resonances associated to the passive cavity created by introducing the filter. Note that we discounted the noise peak at 100 GHz since it was due to other nonlinear effects, and was not an exact integer sub-multiple of the 400 GHz peak (exactly measured at 404 GHz). The capability of detecting these subtle noise features demonstrates the potential and utility of our RF spectrum analyzer.

## 6. Discussion

There are two fundamental physical processes that affect the performance of this device. First, waveguide dispersion imposes a maximum intrinsic bandwidth limitation. Second, the nonlinear efficiency of the device limits its overall sensitivity. From Eq. (6) we have:

$$\Delta\varphi \approx L \cdot \left(\beta^{(1)}(\omega_0) - \beta^{(1)}(\omega_S)\right)\Delta\omega_S \qquad (9)$$

where $\Delta\varphi$ is the phase mismatch and $L$ is the propagation length inside the waveguide. From Eq. (9) one can obtain the maximum walk-off bandwidth $\Delta f_{Max}=\Delta\omega_S/2\pi$ by considering $L = L_{WG}$ and $\Delta\varphi=\pi$, where $L_{WG}$ is the waveguide length. The equation for $\Delta f_{Max}$ can be further simplified by introducing the dispersion parameter $D$, accounting for the dependence of $\beta^{(1)}$ on $\lambda$ and approximating a Taylor series, which leads to:

$$\left(\beta^{(1)}(\omega_0) - \beta^{(1)}(\omega_S)\right) \approx \left| D \cdot \Delta\lambda + \frac{1}{2} \cdot \left.\frac{\partial D}{\partial \lambda}\right|_{\lambda_S} \cdot \Delta\lambda^2 \right| \qquad (10)$$

where Δλ corresponds to the wavelength distance between the signal carrier and the CW seed. Combining Eq. (9) and Eq. (10) yields the following expression for the device bandwidth:

$$\Delta f_{Max} = \frac{1}{L_{WG}\left|2D \cdot \Delta\lambda + \left.\frac{\partial D}{\partial \lambda}\right|_{\lambda_S} \cdot \Delta\lambda^2\right|} \qquad (11)$$

Equation (11) is equivalent to that reported in [13] and used in [14, 15] for the evaluation of the intrinsic bandwidth of the RF spectrometer. For our device the calculated intrinsic bandwidth at 3dB is $\Delta f_{Max}$=130.2 THz. This value has been obtained by plugging the following values into Eq. (11): $\Delta\lambda$=20nm; $D$=2.5ps·nm$^{-1}$·Km$^{-1}$; $\partial D/\partial\lambda|_{\lambda_S}$=-0.23 ps·nm$^{-2}$·Km$^{-1}$ ; and $L_{WG}$=4cm.

The maximum $\Delta f_{3dB}$ practically achievable would likely be significantly reduced by considering higher order dispersion terms not included in [13], and possibly other effects. Nevertheless, on one hand Eq. (11) can be used as a sort of FOM to compare our RF system with the other already published. On the other hand, the extremely large value of the intrinsic bandwidth leaves room for the device optimization via dispersion engineering of the nonlinear waveguides and by equipment upgrade. This would significantly make up for its lower nonlinearity relative to chalcogenides or silicon [13, 14]. Another advantage of our device is that it has negligible TPA ($\alpha^{(2)}$) in the telecom band (even up to 25 GW/cm$^2$ [23]), whereas chalcogenide glass and silicon in particular [28, 29], have much higher TPA and hence a much lower nonlinear FOM (= $n_2/\alpha^{(2)}\lambda$). Our device would therefore allow much higher probe powers, significantly increasing its sensitivity since XPM scales quadratically with power.

## 7. Conclusions

We report an ultra-broadband all-optical RF spectrum analyzer, based on a ~ 4 cm long waveguide, that exhibits a 3dB bandwidth greater than 2.5 THz, and possibly much higher since simple theoretical estimates of the walk-off bandwidth yield 130.2 THz. We use this device to characterize mode-locked lasers with sub-picosecond pulses at repetition rates of up to 400 GHz, and uncover new noise related behaviors not observable with other methods. Our CMOS-compatible device offers the promise of being a key tool for monitoring and characterizing ultra-high speed devices and photonic networks.

## Acknowledgements

R.M would like to acknowledge support from NSERC (The National Science and Engineering Research Council) through the Strategic Grant Program. M.F. acknowledges support from the People Program (Marie Curie Actions) International Outgoing Fellowship (ATOMIC) under REA grant agreement n° [329346]. C.R. acknowledges financial support through a Vanier Canada Graduate Scholarship. AP acknowledges support from the People Program (Marie Curie Actions) Incoming International Fellowship (CHRONOS) under REA grant agreement n° [327627]. M.P. acknowledges support from the People Programme (Marie Curie Actions) Career Integration Grant (THEIA) under REA grant agreement n° 630833. M.C. acknowledges support from the IOF People Programme (Marie Curie Actions) of the European Union's Seventh Framework Programme FP7-2012 (KOHERENT) under REA grant agreement n° [299522].